# Up to 40 % reduction of the GaAs band gap energy via strain engineering in core/shell nanowires

L. Balaghi[1,2], G. Bussone[3], R. Grifone[3], R. Hübner[1], J. Grenzer[1], M. Ghorbani-Asl[1], A. Krasheninnikov[1], H. Schneider[1], M. Helm[1,2], E. Dimakis[1]

[1] Institute of Ion Beam Physics and Materials Research, Helmholtz-Zentrum Dresden-Rossendorf, 01328 Dresden, Germany

[2] Center for Advancing Electronics Dresden (cfaed), Technische Universität Dresden, 01062 Dresden, Germany

[3] PETRA III, Deutsches Elektronen-Synchrotron (DESY), 22607 Hamburg, Germany

The great possibilities for strain engineering in core/shell nanowires have been explored as an alternative route to tailor the properties of binary III-V semiconductors without changing their chemical composition. In particular, we demonstrate that the GaAs core in GaAs/In$_x$Ga$_{1-x}$As or GaAs/In$_x$Al$_{1-x}$As core/shell nanowires can sustain unusually large misfit strains that would have been impossible in conventional thin-film heterostructures. The built-in strain in the core can be regulated via the composition and the thickness of the shell. Thick enough shells become almost strain-free, whereas the thin core undergoes a predominantly-hydrostatic tensile strain, which causes the reduction of the GaAs band gap energy. For the highest strain of 7 % in this work (obtained for x=0.54), a remarkable reduction of the band gap by 40 % was achieved in agreement with theoretical calculations. Such strong modulation of its electronic properties renders GaAs suitable for near-infrared nano-photonics and presumably high electron mobility nano-transistors.

Introduction

III-V compound semiconductors have fueled many breakthroughs in physics and technology owing to their direct band gap and high electron mobility. It has also been very important that these fundamental properties can be tailored in (qua)ternary alloys by selecting the chemical composition appropriately. $In_xGa_{1-x}As$ is a representative example, where In-content x can be chosen to provide an appropriate band gap energy for near-infrared photonics and a low effective mass of electrons for high-mobility transistors. More recently, III-V semiconductors in the form of free-standing nanowires (NWs) have found new strengths for a wide range of future applications in nanotechnology, i.e. solar cells with enhanced light-absorption, lasers with sub-wavelength size, junction-less or tunnel field effect transistors, single-photon emitters, thermoelectrics, chemical sensors, etc. Owing to their small footprint, NWs can be also grown epitaxially without dislocations on lattice-mismatched substrates, enabling the monolithic integration of dissimilar materials with complementary properties, like III-V semiconductors and Si or graphene.

A distinct feature of the NW geometry is the high surface-to-volume ratio, which offers the opportunity to create core/shell heterostructures of highly-mismatched materials well beyond the limits for coherent growth in equivalent thin-film heterostructures. The misfit can be accommodated via elastic deformation of not only the shell, but also the core, depending on the relative thicknesses and chemical compositions [1, 2]. This increases the capabilities for engineering the strain and, thus, the band profile and the electronic properties of the heterostructure [3–8]. Compared to quantum dot heterostructures, where elastic accommodation of large misfit stresses is also possible [9], the hetero-interface in NWs can be several micrometers long, allowing for practical use in a wide variety of device concepts, i.e. thermoelectrics, photovoltaics, lasers and electronics [10].

Here, we have investigated the strain in highly-mismatched $GaAs/In_xGa_{1-x}As$ and $GaAs/In_xAl_{1-x}As$ core/shell NWs, and its effect on the electronic properties of the GaAs core. Specifically, our work shows how to surmount strain-induced difficulties in the growth, how the misfit strain is distributed between the core and the shell depending on the design of the heterostructure and, most important, how to obtain highly-strained cores with a sizeable change in their electronic properties. Although our findings could be extended to other materials as well, they are of particular importance to GaAs because its electronic properties can now be tailored to fit to photonics for optical fiber telecommunications or to high-speed electronics. The use of a binary alloy instead of a ternary one (e.g. $In_xGa_{1-x}As$) would be advantageous because phenomena like phase separation, surface segregation or alloy disorder that typically exist in (qua)ternary alloys and limit the performance of photonic or electronic devices, become now irrelevant.

Results

Growth of strained core/shell nanowires

Vertical GaAs/In$_x$Ga$_{1-x}$As and GaAs/In$_x$Al$_{1-x}$As core/shell NWs were grown on Si(111) substrates in the self-catalyzed mode by molecular beam epitaxy (MBE). The GaAs core NWs had a diameter of 20-25 nm and a length of 2 µm, whereas the shell thickness (L$_S$) and composition (x) were varied independently according to the needs of our study. Figures 1a and 1b depict side-view scanning electron microscopy (SEM) images of bare GaAs core NWs (without shell) and GaAs/In$_x$Ga$_{1-x}$As core/shell NWs (x=0.2, L$_S$=40 nm), respectively. The growth conditions for the shell were tuned to obtain homogeneous thickness and composition around the core NWs. Specifically, the growth of the shell was performed at a considerably low substrate temperature (370 °C) with a continuous substrate-rotation of 20 rpm. The structure and composition of the NWs were evaluated with transmission electron microscopy (TEM). Figures 1c and 1d show elemental maps along and perpendicular to the NW axis as measured with energy-dispersive X-ray (EDX) spectroscopy. The incorporation of In into the shell was found very homogeneous except for the NW corners, where the incorporation was reduced, giving rise to six $<11\bar{2}>$ lines of lower x (the same occurs in In$_x$Al$_{1-x}$As shells). This phenomenon has also been observed by others and attributed to a lower sticking of In adatoms on the $(11\bar{2})$ nano-facets. The shell adopts the crystal structure of the core, i.e. both the core and the shell grow in the zinc blende structure and only the two ends of the NWs contain rotational twin (111) planes that form in the beginning and the end of the GaAs growth (owing to transient changes of the droplet contact angle). Finally, the coherent growth along the $[11\bar{2}]$ crystallographic direction of the core/shell interface is evidenced by the absence of misfit dislocations as shown in Figure 1e (high resolution TEM image of the region shown with the yellow square in Figure 1d).

The large core/shell misfit (f) for higher x, however, can have implications on the NW growth. The cross sectional EDX elemental map of a GaAs/In$_x$Al$_{1-x}$As NW with x=0.44 and L$_S$=80 nm in Figure 1f shows curved lines of lower x. This is understood as follows. Initially, the shell grows preferentially on one side of the NW core, as suggested by Day et al. [5] for highly strained Si/Ge core/shell NWs, imposing a higher misfit stress on that side of the core. As a result, the NWs bend towards the thinner-shell side [5, 6, 8, 11]. As the shell grows thicker, though, the stress asymmetry is extinguished and the NWs become straight again. The transient bending of NWs in the beginning of the shell growth was also observed *in situ* using reflection high energy electron diffraction (RHEED), where the diffraction pattern developed powder-like features for a short period of time.

Analysis of strain in core/shell nanowires

The strain in core/shell NWs was measured by micro-Raman scattering spectroscopy at 300 K (using a frequency doubled Nd:YAG laser with λ=532 nm). The measurements were performed in $\bar{x}(z,z)x$ back scattering configuration on single NWs, which had been transferred previously on an Au-coated Si substrate. An example of a GaAs/In$_x$Ga$_{1-x}$As core/shell NW with x=0.2 and L$_S$=40 nm is shown in Figure 2a in comparison with a bare GaAs NW (without shell). The spectrum of the bare GaAs NW is dominated by scattering from transverse optical (TO) phonons, with the peak position being in perfect agreement with that for strain-free bulk GaAs (longitudinal optical (LO) phonon transitions are forbidden in the particular measurement geometry, but a weak signal is still present). In contrast, the spectrum of the core/shell NW shows a more complex structure. Using Lorentzian curves for the fitting of the line shape, we identified three scattering contributions, i.e. scattering in the core from GaAs TO phonons and scattering in the shell from GaAs-like and InAs-like TO phonons. Measuring the relative peak shift ($\Delta\omega/\omega$) with respect to the strain-free position for GaAs TO and GaAs-like TO phonons, we deduced the amount of hydrostatic strain $\Delta V/V$ in the core and in the shell, respectively, using the following equation:

$$\Delta V/V = 1/\gamma \cdot \Delta\omega/\omega$$

, where $\gamma$ is the Grüneisen parameter of bulk GaAs or bulk $In_xGa_{1-x}As$; homogeneously strained core and shell have been assumed in our analysis.

Figure 2b shows the Raman shifts and the corresponding strain in $GaAs/In_xGa_{1-x}As$ core/shell NWs as a function of $L_S$ for x=0.20. For the smallest $L_S$ both the shell and the core are strained, i.e. the shell is compressively strained, whereas the core is tensile-strained. With increasing $L_S$, though, the shell becomes less strained and the core more strained. In other words, the compressive misfit strain that exists in thin shells is elastically relaxed with increasing $L_S$ by stretching the core (later we will show that this is not the only mechanism of strain relaxation in the shell). Eventually, for $L_S \geq 40$ nm the shell becomes almost strain-free, whereas the strain in the core saturates at 3.2 %. These results already show that thin enough NWs can be used as flexible substrates for overgrowth with lattice mismatched shells, going far beyond what is possible in equivalent planar heterostructures.

The large strain in the GaAs core was verified for a selected number of samples using high-resolution X-ray diffraction (XRD) from synchrotron light sources. The lattice constants were measured along three orthogonal crystallographic directions, namely the $[111]$ (parallel to the NW axis), the $[11\bar{2}]$ (perpendicular to the NW axis and parallel to the core/shell interface) and the $[1\bar{1}0]$ (perpendicular to the NW axis and normal to the core/shell interface) directions. For this purpose, three-dimensional reciprocal space maps were recorded around the $(111)$, $(20\bar{2})$ and $(22\bar{4})$ reflections of NW ensembles with x=0.20 and $L_S$ = 0, 5, 10, 40 and 80 nm. As an example, the projection of the $(111)$ reciprocal space map on the $(q_z, q_x)$ plane for $L_S$= 40 nm is depicted in Figure 2c. The extracted lattice constants of the core, one along ($\alpha_\parallel$) and two perpendicular ($\alpha_\perp$) to the NW axis, are plotted in Figure 2d as a function of $L_S$. The fact that all three lattice constants increase with $L_S$ is a manifestation of the hydrostatic character of strain in the core. Furthermore, $\alpha_\parallel$ increases gradually with $L_S$ from the value of strain-free GaAs to that of strain-free $In_{0.20}Ga_{0.80}As$, and the same happens with the shell (not shown here). This means that the compressive misfit strain in the shell along the $[111]$ direction can be fully relaxed exclusively by stretching the core. On the other hand, $\alpha_\perp$ in the core does not reach the value of strain-free $In_{0.20}Ga_{0.80}As$ with increasing $L_S$, which indicates that the strain in the core perpendicular to the NW axis is less than that along the NW axis.

The strain components along ($\varepsilon_\parallel$) and perpendicular ($\varepsilon_\perp$) to the NW axis were calculated for the GaAs core as $\varepsilon_\parallel = (\alpha_\parallel - \alpha_0)/\alpha_0$ and $\varepsilon_\perp = (\alpha_\perp - \alpha_0)/\alpha_0$ (where $\alpha_0$ is the lattice constant of strain-free GaAs), whereas the corresponding hydrostatic strain was calculated as $\varepsilon_h = \varepsilon_\parallel + 2 \cdot \varepsilon_\perp$ ($\varepsilon_\perp$ is almost the same for $[1\bar{1}0]$ and $[11\bar{2}]$). As shown in the lower plot of Figure 2b, the results for $\varepsilon_h$ (star symbols) are in good agreement with the strain measured by Raman scattering. Thus, the unusually large strain in the GaAs core is verified by two independent experimental techniques.

The amount of tensile strain in the GaAs core depends also on the core/shell misfit (f) or, in other words, the shell composition x. Figure 3 shows the Raman shift and the corresponding strain in GaAs/In$_x$Ga$_{1-x}$As core/shell NWs with different values of x, from 0.10 to 0.55, and L$_S$ = 40 - 80 nm (the nominal values of x were confirmed for a selected number of samples by EDX-TEM). The tensile strain in the core was found to increase linearly with x, whereas the shell remains approximately strain-free. For the highest misfit in this work, i.e. f= 4 % for x=0.55, the tensile strain in the core reaches the remarkably large value of 7 %. We point out that it would have been impossible to grow such highly lattice-mismatched heterostructures in conventional thin-film geometry without forming dislocations. The tensile strain in the core was also measured with Raman spectroscopy for GaAs/In$_x$Al$_{1-x}$As NWs with different x and L$_S$ = 80 nm. The results (open symbols in the lower plot of Figure 3) are similar to those for GaAs/In$_x$Ga$_{1-x}$As NWs because of the similar lattice constants of In$_x$Ga$_{1-x}$As and In$_x$Al$_{1-x}$As for the same x.

Effect of strain on the electronic properties of GaAs core

The effect of strain on the band gap of GaAs core was studied by means of photoluminescence (PL) spectroscopy. The existence of tensile strain with hydrostatic character in the core is expected to reduce the band gap energy. In fact, the band gap energy of the tensile-strained GaAs core and the strain-free In$_x$Ga$_{1-x}$As shell are expected to be similar, which makes their distinction in optical spectra difficult. To avoid any ambiguities, we used GaAs/In$_x$Al$_{1-x}$As core/shell NWs, where the larger band gap energy of strain-free In$_x$Al$_{1-x}$As (larger than 1.38 eV at 12 K for x ≤ 0.54) cannot be confused with that of the tensile-strained GaAs. PL measurements were performed at 12 K (laser excitation at 532 nm) on ensembles of GaAs/In$_x$Al$_{1-x}$As NWs, which had been transferred previously on amorphized Ge substrates (to quench the photoluminescence from Ge). The spectra for different values of x are plotted in Figure 4a. Emission was obtained only in the 0.9 - 1.1 eV range, which is suggestive of radiative recombination of electron-hole pairs inside the tensile-strained GaAs core. The emission peak shifts to lower energies with increasing x, manifesting the effect of the increasing tensile strain in the core. The origin of the second peak that appeared at a higher energy (shifted by 40 meV) for x=0.54 is unclear, but could be possibly associated either with valence band splitting at the Γ point of the Brillouin zone or with strain and/or compositional inhomogeneity in these highly-strained NWs. Figure 4b shows the peak energy ($E_{PL}$) as a function of x and the corresponding strain ($\frac{\Delta V}{V}$) in the core. The linear dependence of $E_{PL}$ on $\frac{\Delta V}{V}$ can be fitted with an equation of the form:

$$E_{PL} = E^o + a \cdot \frac{\Delta V}{V}$$

, where the interception $E^o$ is the strain-free value of $E_{PL}$ and the slope $a$ is the hydrostatic deformation potential of $E_{PL}$. Both fitting parameters ($E^o$= 1.54 eV and $a$= -8.91 eV) were found to be in good agreement with the strain-free band gap energy and the hydrostatic deformation potential, respectively, of bulk GaAs [12–14)]. Thus, $E_{PL}$ can be attributed to the band gap energy of the tensile-strained GaAs core. For the highest strain (obtained for x=0.54), the band gap energy of GaAs at 12 K is reduced from the strain-free value of 1.52 eV to 0.92 eV, i.e. a striking reduction by almost 40 %. This is particularly important for applications in optical fiber telecommunications because the emission from strained GaAs NWs can now cover the O-band and potentially the S-band of telecommunication wavelengths.

If we assume that the strain in GaAs is purely hydrostatic, the corresponding effective mass of electrons $m_e^*$ can be calculated using the following pressure coefficients [15]:

$$\frac{dE_g}{dP} = 12.02 \text{ eV/Mbar}$$

$$\frac{1}{m_e^*}\frac{dm_e^*}{dP} = 6.8 \text{ Mbar}^{-1}$$

, where $dE_g$ is the shift of the band gap energy, which has been measured by PL, induced by a relative pressure $dP$. The results are shown in Figure 4b. As expected, $m_e^*$ in GaAs is reduced with increasing the In-content in the shell, reaching a value of $m_e^*$=$0.0417m_0$ for the highest In-content here, equivalent to a reduction of 38 % from the strain-free value ($0.067m_0$ [14]). In fact, this lowest value of $m_e^*$ is comparable to that for bulk $In_xGa_{1-x}As$ with x= 0.52 – 0.54 that is typically used in high electron mobility transistors (HEMTs) on lattice-matched InP substrates. This means that high-mobility transistors could now be possible with strained GaAs NWs and without the need for lattice-matched substrates.

Conclusion

In conclusion, our results show that the GaAs core in GaAs/$In_xGa_{1-x}As$ or GaAs/$In_xAl_{1-x}As$ core/shell NWs can sustain unusually large misfit strains that would have been impossible in equivalent thin-film heterostructures. The strain of GaAs is tensile, can be regulated via the shell thickness and composition, and exhibits a predominantly hydrostatic character that is similar only to quantum-dot heterostructures. As a result, the electronic properties of GaAs can be strongly modified, as if we had changed its chemical composition by adding In. For the highest strain in this work, a 40 % smaller band gap was obtained and a 38 % smaller effective mass of electrons is presumed. We believe that the possibility to resemble the electronic properties of $In_xGa_{1-x}As$ thin-films with strained GaAs in core/shell NWs will set new design rules for nano-devices. GaAs could replace $In_xGa_{1-x}As$ in the active region of a high-speed transistor or a near-infrared laser and phenomena like phase separation, surface segregation or alloy disorder, which typically exist in ternary alloys and limit the device performance, would become irrelevant. Finally, we anticipate that the concept of using core/shell NWs to tailor the properties of binary alloys can be also adopted for other III-V material systems.

Acknowledgements

The authors express their gratitude for the granted beam-time and financial support at Diamond light source (beamline I07; allocated beam-time: SI15923), as well as for the granted beam-time at the German Synchrotron DESY, PETRA III (beamline P08; allocated beam-time: I-20160337). In particular, the authors thank the beamline staff for their support during the experiments. The authors also thank Joachim Wagner for the technical maintenance of the molecular beam epitaxy laboratory and Annette Kunz for the preparation of TEM specimens at HZDR. Support by the Structural Characterization Facilities Rossendorf at Ion Beam Center is gratefully acknowledged.

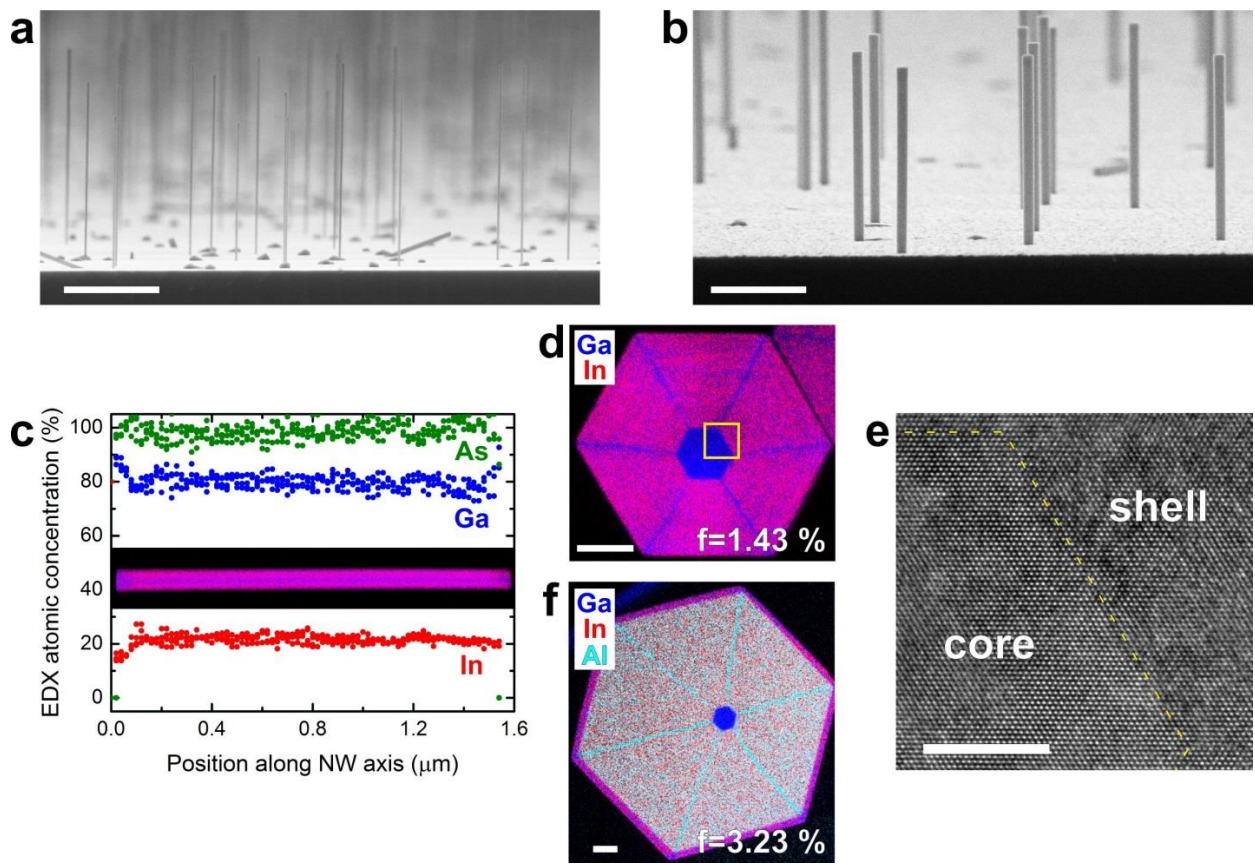

Figure 1. Morphological and compositional analysis of GaAs/In$_x$Ga$_{1-x}$As and GaAs/In$_x$Al$_{1-x}$As core/shell NWs grown on Si(111) substrates. (a) Side-view SEM image of as-grown bare GaAs core NWs and (b) GaAs/In$_x$Ga$_{1-x}$As core/shell NWs (x=0.2, L$_S$=40 nm). (c) EDX compositional analysis along and (d) perpendicular to the axis of one NW from the sample shown in (b). (e) High-resolution TEM of the core/shell interface region shown in (d) with a yellow square. (f) EDX compositional analysis perpendicular to the axis of a GaAs/In$_x$Al$_{1-x}$As core/shell NW (x=0.44, L$_S$=80 nm). The scale bars correspond to 1 μm in (a) and (b), 30 nm in (d) and (f), and 5 nm in (e).

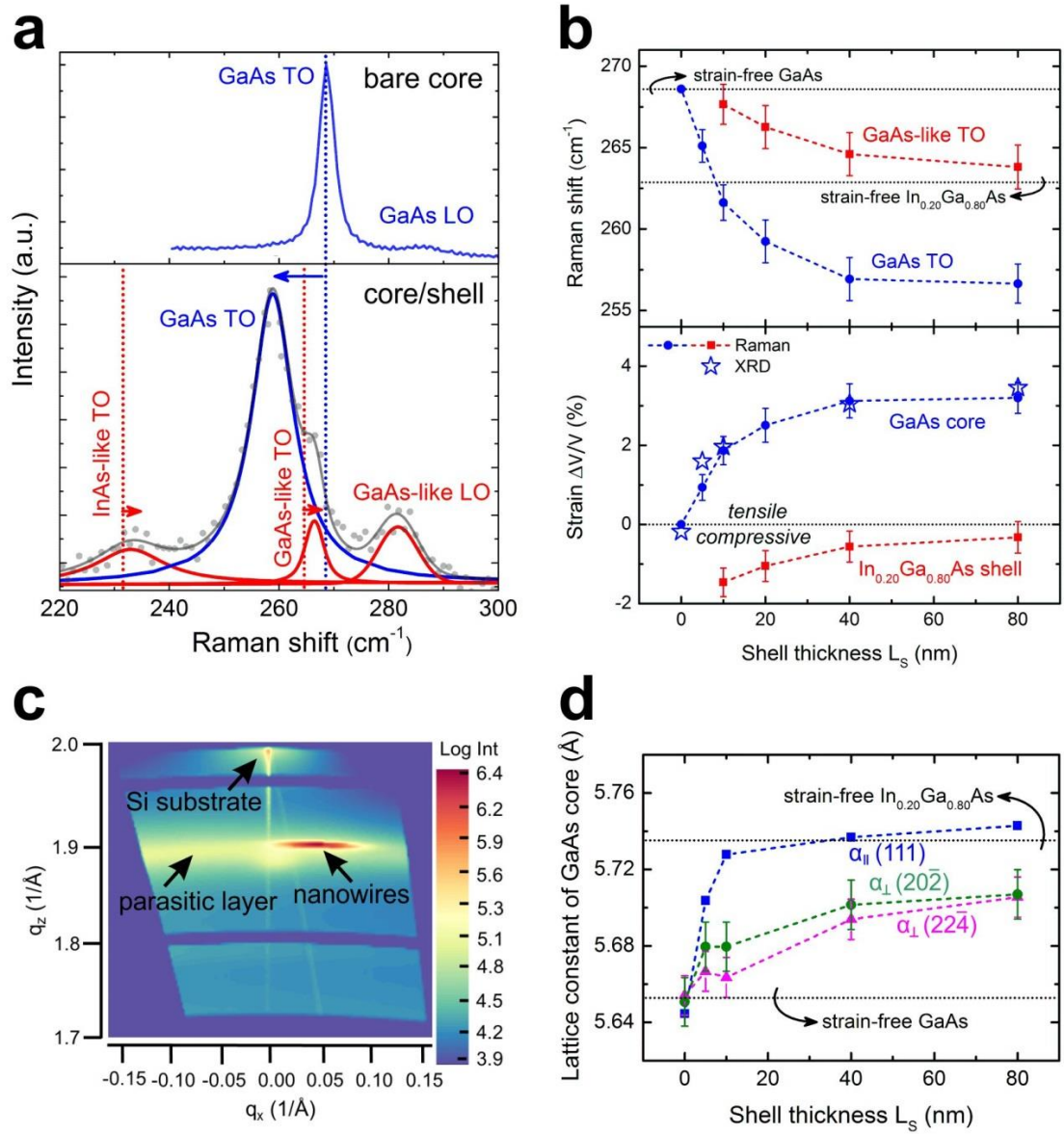

Figure 2. Strain analysis of GaAs/In$_{0.20}$Ga$_{0.80}$As core/shell NWs as a function of shell thickness L$_S$. (a) Raman scattering spectrum at 300 K from a single core NW without shell (upper plot) and from a single core/shell NW with L$_S$=40 nm (lower plot). Phonon peaks in blue are attributed to the core, whereas the ones in red to the shell. (b) Upper plot: Raman shift of the GaAs (core) and the GaAs-like (shell) TO phonons as a function of L$_S$. Lower plot: The corresponding strain as a function of L$_S$. (c) 2D reciprocal space map of the (111) reflection as measured on an ensemble of as-grown NWs with L$_S$=40 nm. (d) XRD-measured lattice constants of the core as a function of L$_S$. $\alpha_\parallel$ is the lattice constant parallel to the NW axis, extracted from the (111) reflection. $\alpha_\perp$ are two orthogonal lattice constants perpendicular to the NW axis, extracted from the $(20\bar{2})$ and the $(22\bar{4})$ reflections.

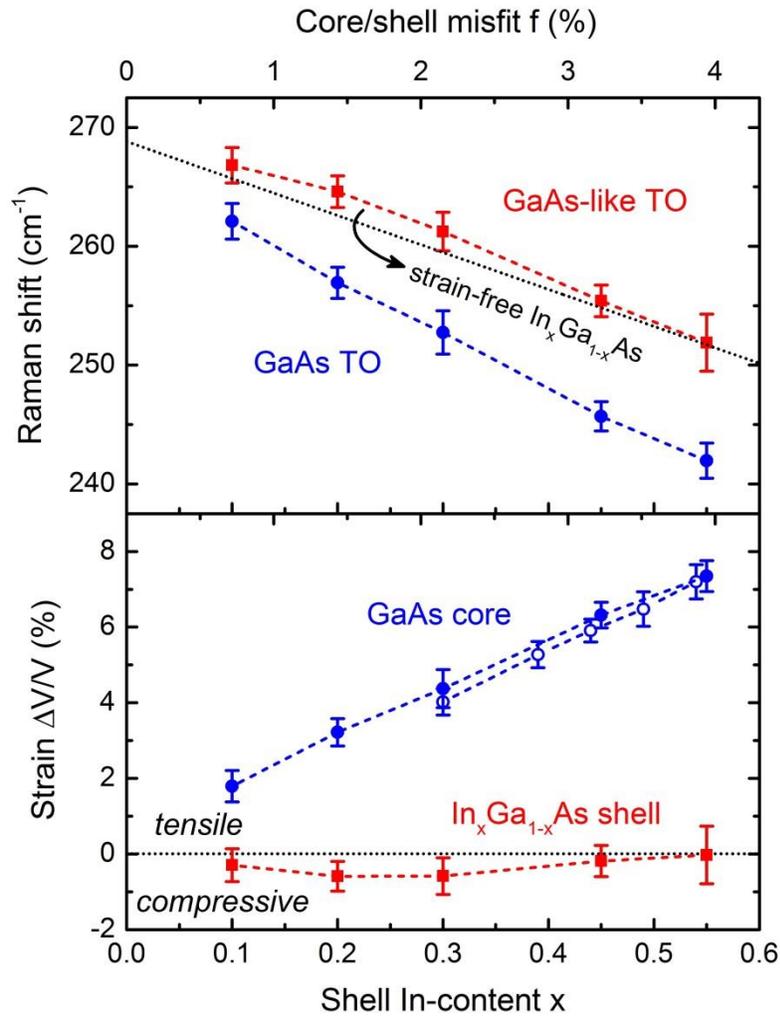

Figure 3. Strain analysis of GaAs/In$_x$Ga$_{1-x}$As core/shell NWs with L$_S$= 40 – 80 nm as a function of In-content x in the shell. Upper plot: Raman shift of the GaAs (core) and the GaAs-like (shell) TO phonons as a function of x. Lower plot: The corresponding strain as a function of x. The open symbols correspond to the strain in the core of GaAs/In$_x$Al$_{1-x}$As core/shell NWs with L$_S$=80 nm.

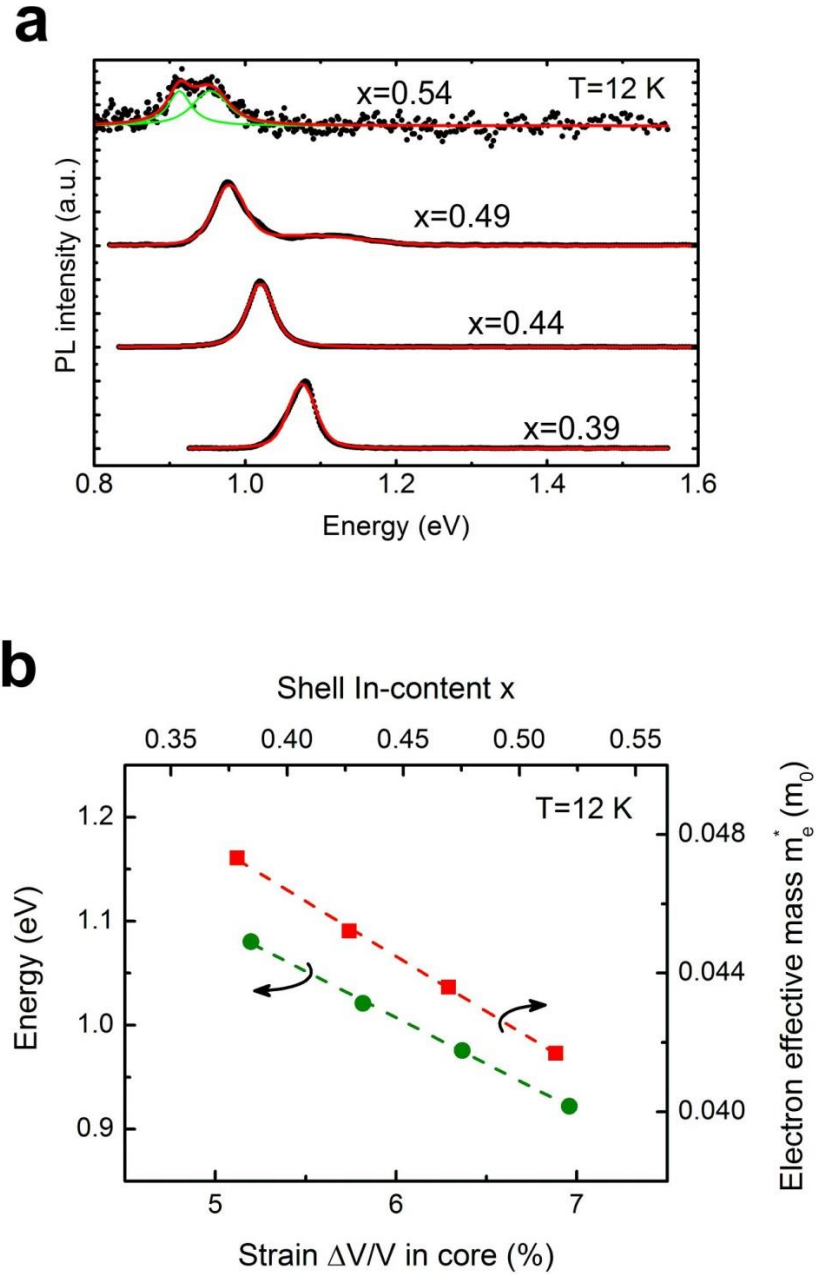

Figure 4. Effect of strain on the band gap energy of the core in GaAs/In$_x$Al$_{1-x}$As core/shell NWs with L$_S$=80 nm. (a) PL spectra as measured on ensembles of NWs with different x at 12 K. (b) PL peak emission at 12 K (green data points) as a function of strain in the core and the corresponding In-content x in the shell. The corresponding calculated electron effective mass is also plotted (red data points).